\documentclass[epj]{svjour}

\pdfoutput=1

\usepackage[T1]{fontenc}
\usepackage{graphicx}
\usepackage{float}
\usepackage{booktabs}
\usepackage{siunitx}
\usepackage{placeins}
\usepackage{subcaption}
\usepackage{amsmath}
\usepackage{amssymb}
\usepackage{comment}
\usepackage{hyperref}
\usepackage{lmodern}
\usepackage[dvipsnames]{xcolor}
\usepackage{pgfgantt}
\usepackage{tabularx}
\usepackage{array}
\title{Constraining Tsallis Corrections to Photon Reheating from Electron-Positron Annihilation: A Phenomenological Approach}

\author{Matias P. Gonzalez}

\institute{
Departamento de F\'{\i}sica, Universidad Cat\'olica del Norte,
Avenida Angamos 0610, Chile\\
\email{matias.gonzalez03@alumnos.ucn.cl}
}

\date{Received: date / Revised version: date}
\abstract{In this work we generalize the entropy transfer from electron-positron
annihilation to photons in the early Universe. The generalization is
implemented within the Tsallis formalism by using generalized distribution functions derived from Curado-Tsallis constraints. Through this deformation, the entropy density of the electromagnetic sector is modified, while the photon component is kept extensive. Therefore, the nonextensive correction is introduced only in the $e^-e^+$ pairs. This affects the entropic degrees of freedom before electron-positron annihilation and consequently modifies the temperature ratio $T_\nu/T_\gamma$. The resulting correction is then mapped into an effective value of \(N_{\rm eff}\) within the instantaneous decoupling approximation. Comparing this effective thermodynamic estimate with CMB$+$BAO data, and using BBN only as an illustrative Gaussian comparison, we obtain an order-of-magnitude phenomenological interval for the nonextensive parameter. The result should not be interpreted as a precision neutrino decoupling constraint, since a full treatment would require solving the kinetic evolution of the neutrino sector, including non-instantaneous decoupling, finite-temperature QED effects and possible spectral distortions.
}

\begin{document}

\maketitle

\section{Introduction}
\label{sec:introduction}

The thermal history of the early Universe provides one of the most precise connections between microscopic particle physics and cosmological observables. At
early times, the Universe was composed by a hot and dense plasma whose macroscopic evolution is well described, in the standard picture, by Boltzmann-Gibbs equilibrium thermodynamics embedded in an expanding FLRW background (homogeneous and isotropic) \cite{KolbTurner1990,Dodelson2003,Weinberg2008}.
In the radiation dominated era, the content of particles of the plasma is
encoded in the effective energy degrees of freedom and the entropy degrees of freedom,
$g_*(T)$ and $g_{*s}(T)$, which determine both the expansion rate through the Hubble parameter $H$ and the thermal evolution of the different species \cite{LesgourguesManganoMielePastor2013,Husdal2016}. This framework is particularly relevant around the MeV era, where neutrino decoupling and electron-positron pair annihilation set the initial conditions for subsequent cosmological observables.

In the standard thermal history, neutrinos remain coupled to the
electromagnetic plasma through weak interactions until their interaction rate
falls below the Hubble expansion rate. After decoupling, neutrinos free stream
and their temperature redshifts approximately as $T_\nu\propto a^{-1}$ where $a$ is the scale factor, while
the electromagnetic plasma continues to exchange entropy between its components. The
annihilation of electron-positron pairs then transfers entropy to photons,
reheating the photon bath relative to the neutrino background. In the
instantaneous decoupling approximation this gives the well known ratio
$(T_\nu/T_\gamma)_{\rm std}=(4/11)^{1/3}$, which enters directly in the definition of the effective number of relativistic species,
$N_{\rm eff}$ \cite{KolbTurner1990,Dodelson2003,LesgourguesManganoMielePastor2013}.
Precision treatments go beyond this approximation by including
non-instantaneous neutrino decoupling, flavor oscillations, finite temperature
QED corrections, and kinetic effects in the Boltzmann equations
\cite{ManganoEtAl2005,DeSalasPastor2016,AkitaYamaguchi2020,FrousteyPitrouVolpe2020,BennettEtAl2021Neff}. Related analyses have also examined
the freeze-out of electron-positron annihilation and its role in the MeV
thermal plasma \cite{ThomasDezenGrohsKishimoto2020}.

The effective number of relativistic species is therefore a sensitive probe of the thermal evolution of the primordial plasma. Observationally,
$N_{\rm eff}$ is constrained by the cosmic microwave background and by
large-scale structure information, with Planck 2018 combined with BAO
providing a standard reference dataset \cite{Planck2018CosmoParams}. Primordial abundance measurements provide a complementary probe through BBN, especially
through the deuterium abundance ${\rm D/H}$ and the helium mass fraction $Y_p$ \cite{CyburtFieldsOliveYeh2016,PitrouCocUzanVangioni2018}. Recent and
future CMB analyses can further sharpen the allowed window for extra
radiation or non-standard thermal histories \cite{ACTDR6Extended2025,GoldsteinHill2026Neff}. The high precision of the CMB blackbody spectrum also
motivates keeping the photon sector standard in minimal phenomenological
extensions of the electromagnetic plasma \cite{Fixsen2009}.

A possible way to study controlled departures from the standard
Boltzmann-Gibbs description is provided by Tsallis nonextensive statistics
\cite{Tsallis1988,Tsallis2009Book}. In this framework, the entropy functional depends on a real parameter $q$, with the extensive Boltzmann-Gibbs limit
recovered when $q\to1$. The formalism has been used to describe systems with long range interactions, correlations, memory effects, or quasi-stationary
behavior, and the associated generalized distributions can produce either depleted or enhanced high energy tails depending on the sign of $q-1$ \cite{CuradoTsallis1991,TsallisMendesPlastino1998,LimaSilvaPlastino2001}. Generalized quantum distributions and related
applications in relativistic or high-energy systems have also been discussed in several contexts \cite{BuyukilicDemirhanGulec1995,TirnakliBuyukilicDemirhan1998,TorresTirnakli1998,Mitra2018,CleymansWorku2012}. In cosmology, deformations using Tsallis statistics and
bounds on departures from extensivity have been explored in BBN, dark matter, and generalized entropic scenarios \cite{TorresVucetichPlastino1997,PlastinoEtAl2004,GhoshalLambiase2021,JizbaLambiase2023,JizbaEtAl2024,Gonzalez2026NeffTsallis}. More recently, Tsallis-type structures have also been connected with generalized uncertainty relations, coherent states and quantum gravity motivated deformations, suggesting that nonextensive thermostatistics can provide a useful language beyond standard kinetic theory scenarios \cite{JizbaEtAl2022PRD,JizbaEtAl2023PRD}.

In this work we focus on a minimal thermodynamic implementation of this idea. We do not deform the neutrino sector directly. Instead, we introduce a Tsallis deformation in the electron-positron occupation numbers and propagate it into the entropy budget of the electromagnetic plasma. This modifies the electromagnetic entropic degrees of freedom \(g_{*s,{\rm EM},q}(T)\) around the electron-positron annihilation epoch. This species selective prescription should be understood as a minimal benchmark setup rather than as a complete microscopic description of the MeV plasma. The motivation is to isolate the entropy-transfer effect of the annihilating \(e^-e^+\) sector, while keeping photons extensive in order to preserve the standard thermal electromagnetic bath and keeping neutrinos standard so that the induced shift in \(N_{\rm eff}\) arises only through the modified ratio \(T_\nu/T_\gamma\). Within the instantaneous decoupling approximation, the modified entropy transfer changes the neutrino-to-photon temperature ratio, which is then mapped into an effective thermodynamic estimate of \(N_{\rm eff}\). This construction therefore provides a controlled phenomenological benchmark for photon reheating. It should not be interpreted as a full kinetic calculation of neutrino decoupling. In the statistical comparison, CMB$+$BAO is used as the main quantitative reference, while the BBN-inferred value of \(N_{\rm eff}\) is included only as an illustrative Gaussian comparison. A precision treatment would require solving the neutrino Boltzmann equations with collision terms, finite-temperature QED corrections and possible spectral distortions, together with a full BBN abundance likelihood
\cite{PisantiEtAl2008,PisantiManganoMieleMazzella2021}.

The paper is organized as follows. In Sec.~\ref{sec:thermal_history} we review the standard thermal history around neutrino decoupling and
electron-positron annihilation, emphasizing the entropy-conservation argument that leads to the standard neutrino to photon temperature ratio. In
Sec.~\ref{sec:tsallis_statistics} we summarize the Tsallis statistical
framework and the generalized distribution functions used in the analysis. In
Sec.~\ref{sec:thermodynamic_integrals} we derive the thermodynamic integrals for the Tsallis deformed electron-positron plasma. In
Sec.~\ref{sec:entropy_degrees_freedom} we construct the electromagnetic
entropy density and the corresponding entropic degrees of freedom. In
Sec.~\ref{sec:modified_temperature_ratio} we compute the modified
temperature ratio between neutrinos and photons induced by the deformed entropy transfer due to the Tsallis formalism. In Sec.~\ref{sec:mapping_neff} we map this correction into an effective value of \(N_{\rm eff}\). In Sec.~\ref{sec:likelihood_results} we perform a phenomenological comparison with observationally inferred values of \(N_{\rm eff}\) and present the resulting order-of-magnitude interval for \(q\). Finally, Sec.~\ref{sec:conclusions} summarizes the conclusions and discusses possible extensions.

\section{Thermal history around electron-positron annihilation}
\label{sec:thermal_history}

Around the $T\sim~$MeV epoch, the primordial plasma is composed mainly of photons, electrons, positrons, and neutrinos \cite{KolbTurner1990,Dodelson2003}. At temperatures above neutrino decoupling, weak interactions keep the neutrino sector in approximate thermal contact with the electromagnetic plasma. The expansion is radiation dominated, with
\begin{equation}
    \rho_r(T)=\frac{\pi^2}{30}g_*(T)T^4,
    \qquad
    s(T)=\frac{2\pi^2}{45}g_{*s}(T)T^3,
    \label{eq:rho_s_standard}
\end{equation}
and
\begin{equation}
    H^2=\frac{8\pi G}{3}\rho_r .
    \label{eq:friedmann_radiation}
\end{equation}
The effective functions $g_*(T)$ and $g_{*s}(T)$ encode, respectively, the relativistic contributions to the energy and entropy densities \cite{Weinberg2008,Husdal2016}. Neutrino decoupling occurs when the weak interaction rate becomes comparable to the Hubble rate \cite{LesgourguesManganoMielePastor2013},
\begin{equation}
    \Gamma_\nu(T_{\rm dec})\simeq H(T_{\rm dec}),
    \qquad
    \Gamma_\nu\sim G_F^2T^5 ,
    \label{eq:neutrino_decoupling}
\end{equation}
where $G_F$ is the Fermi constant. This gives $T_{\rm dec}$ (temperature of decoupling) of order MeV. After decoupling, neutrinos approximately free stream and their temperature redshifts as $T_\nu\propto a^{-1}$ where $a$ is the scale factor, while the electromagnetic plasma remains thermally coupled.

The annihilation of electron-positron pairs transfers entropy to the photon bath. In the instantaneous decoupling approximation, the comoving entropy of the electromagnetic sector is conserved \cite{KolbTurner1990,LesgourguesManganoMielePastor2013},
\begin{equation}
    s_{\rm EM}a^3
    =
    \frac{2\pi^2}{45}
    g_{*s}^{\rm EM}(T_\gamma)T_\gamma^3a^3
    =
    \mathrm{constant}.
    \label{eq:em_entropy_conservation}
\end{equation}
Before annihilation,
\begin{equation}
    g_{*s}^{\rm EM,before}
    =
    2+\frac{7}{8}(2+2)
    =
    \frac{11}{2},
    \label{eq:gstar_em_before}
\end{equation}
whereas after annihilation only photons remain in the electromagnetic bath,
\begin{equation}
    g_{*s}^{\rm EM,after}=2.
    \label{eq:gstar_em_after}
\end{equation}
Hence,
\begin{equation}
    \left(\frac{T_\nu}{T_\gamma}\right)_{\rm std}
    =
    \left(
    \frac{g_{*s}^{\rm EM,after}}
         {g_{*s}^{\rm EM,before}}
    \right)^{1/3}
    =
    \left(\frac{4}{11}\right)^{1/3}.
    \label{eq:Tnu_Tgamma_standard}
\end{equation}

After $e^-e^+$ annihilation, the radiation density is conventionally written as \cite{LesgourguesManganoMielePastor2013}
\begin{equation}
    \rho_r
    =
    \rho_\gamma
    \left[
    1+
    \frac{7}{8}
    \left(\frac{T_\nu}{T_\gamma}\right)^4
    N_{\rm eff}
    \right],
    \qquad
    \rho_\gamma=\frac{\pi^2}{15}T_\gamma^4 .
    \label{eq:rho_r_neff_general}
\end{equation}
Using Eq.~\eqref{eq:Tnu_Tgamma_standard}, this reduces to
\begin{equation}
    \rho_r
    =
    \rho_\gamma
    \left[
    1+
    \frac{7}{8}
    \left(\frac{4}{11}\right)^{4/3}
    N_{\rm eff}
    \right].
    \label{eq:rho_r_neff_standard}
\end{equation}
A complete Standard Model treatment includes non-instantaneous neutrino decoupling and finite-temperature QED effects \cite{ManganoEtAl2005}, which shift the theoretical prediction to $N_{\rm eff}^{\rm std}\simeq 3.044$ \cite{DeSalasPastor2016,AkitaYamaguchi2020,FrousteyPitrouVolpe2020,BennettEtAl2021Neff}. This standard picture fixes the baseline thermal evolution. The nonextensive extension introduced below modifies the electron-positron distribution functions entering the thermodynamic integrals. In the present approximation, we propagate this modification only into the electromagnetic entropy budget and into the instantaneous decoupling estimate of \(T_\nu/T_\gamma\). The corresponding value of \(N_{\rm eff}\) should therefore be understood as an effective thermodynamic mapping, not as the result of a full kinetic neutrino decoupling treatment.

\section{Tsallis statistics and nonextensive distributions}
\label{sec:tsallis_statistics}

The statistical framework used in this work is based on the nonextensive thermostatistics introduced by Tsallis as a generalization of the Boltzmann-Gibbs formulation \cite{Tsallis1988}. Its central object is the entropy functional
\begin{equation}
    S_q \equiv k \, \frac{1-\sum_i p_i^q}{q-1},
    \label{eq:tsallis_entropy}
\end{equation}
where $\{p_i\}$ denotes the probability distribution over the microscopic states of the system, $k$ is a positive constant that can be identified with Boltzmann's constant, and $q\in\mathbb{R}$ is the nonextensive parameter. The standard Boltzmann-Gibbs entropy is recovered in the extensive limit,
\begin{equation}
    \lim_{q\to 1} S_q = -k\sum_i p_i \ln p_i \equiv S_{\rm BG}.
    \label{eq:bg_limit}
\end{equation}
Thus, the parameter $q$ controls the departure from the ordinary exponential-statistical structure. In the context considered here, this departure is treated as a phenomenological deformation of thermal distributions during the radiation era. A distinctive feature of Eq.~\eqref{eq:tsallis_entropy} is that it is not additive in the Boltzmann-Gibbs sense. For two statistically independent systems $A$ and $B$, such that $p_{ij}^{A+B}=p_i^A p_j^B$, Tsallis entropy satisfies the pseudo-additive composition law \cite{Tsallis2009Book},
\begin{equation}
    S_q(A+B)
    =
    S_q(A)+S_q(B)
    +
    \frac{1-q}{k}S_q(A)S_q(B).
    \label{eq:tsallis_pseudoadditivity}
\end{equation}
The last term measures the deviation from ordinary additivity and vanishes when $q\to 1$. For $q<1$, the correction is positive and the entropy is super-additive, whereas for $q>1$ it is sub-additive. In applications to systems with correlations, long range interactions, or quasi-stationary behavior, this property provides a compact way of parametrizing deviations from the standard Boltzmann-Gibbs description \cite{LimaSilvaPlastino2001}. In the present cosmological setting, the parameter $q$ is not introduced as a modification of gravity, but as a deformation of the statistical weights entering the thermal distribution functions.

To construct the corresponding equilibrium distributions, we maximize $S_q$ subject to the standard normalization condition and to macroscopic constraints. In this work we use the Curado-Tsallis constraints implemented through unnormalized $q$-expectation values for the particle number and internal energy \cite{CuradoTsallis1991,TsallisMendesPlastino1998},
\begin{equation}
    \sum_i p_i = 1,
    \qquad
    \overline{E}=\sum_i p_i^q E_i,
    \qquad
    \overline{N}=\sum_i p_i^q N_i.
    \label{eq:ct_constraints}
\end{equation}
This choice leads to a simple deformation of the usual Bose-Einstein, Fermi-Dirac, and Maxwell-Boltzmann distributions, which is useful for numerical implementations in early Universe thermodynamics.

With the convention adopted here, the $q$-exponential is written as \cite{Tsallis2009Book}
\begin{equation}
    e_q(x)
    =
    \left[1+(q-1)x\right]^{1/(q-1)},
    \qquad
    \lim_{q\to 1} e_q(x)=e^x,
    \label{eq:q_exponential}
\end{equation}
whenever the quantity inside brackets is positive. The associated generalized occupation number is
\begin{align}
    f_q(E)
    &=
    \frac{1}{
    \left[1+(q-1)\beta(E-\mu)\right]^{1/(q-1)}
    +\xi
    }\nonumber\\
    &=
    \frac{1}{e_q\!\left(\beta(E-\mu)\right)+\xi},
    \qquad
    \beta\equiv \frac{1}{T},
    \label{eq:q_distribution}
\end{align}
where $\xi=-1$ corresponds to Bose-Einstein statistics, $\xi=+1$ to Fermi-Dirac statistics, and $\xi=0$ to Maxwell-Boltzmann statistics \cite{BuyukilicDemirhanGulec1995,TirnakliBuyukilicDemirhan1998,TorresTirnakli1998}. Equation~\eqref{eq:q_distribution} reduces to the standard equilibrium distribution in the limit $q\to 1$. The sign of $q-1$ determines the high energy behavior of the distribution. For $q>1$, the distribution develops a power law high energy tail relative to the Boltzmann-Gibbs case. For $q<1$, the bracket in Eq.~\eqref{eq:q_exponential} imposes a finite-domain condition,
\begin{equation}
    1+(q-1)\beta(E-\mu)>0,
    \label{eq:compact_support_condition}
\end{equation}
which produces an effective cutoff in the accessible range of energies. This distinction is relevant when the deformed distributions are propagated into thermodynamic integrals, since even mild deviations from $q=1$ can modify the relativistic energy density, entropy density, and therefore the effective radiation content of the Universe. In particular, during the radiation era these modifications can be mapped into shifts of quantities such as the effective number of neutrino species, $N_{\rm eff}$, or into changes in the entropy transfer process associated with electron-positron annihilation \cite{LesgourguesManganoMielePastor2013,Gonzalez2026NeffTsallis}. Thus, Eq.~\eqref{eq:q_distribution} provides the statistical input from which the thermodynamic consequences studied in this work are computed.

\section{Relevant thermodynamic integrals}
\label{sec:thermodynamic_integrals}

Once the generalized distribution functions have been specified, the next step is to propagate them into the macroscopic thermodynamic quantities. The relevant observables for the present analysis are the energy density and pressure, obtained from the phase-space moments of the Tsallis distribution function. For a particle species with internal degeneracy $g$, dispersion relation $E(p)=\sqrt{p^2+m^2}$, and occupation number $f_q(E)$, we define \cite{KolbTurner1990,Dodelson2003}
\begin{align}
    \rho_q
    &=
    g\int \frac{d^3p}{(2\pi)^3}\,
    E(p)\,f_q(E),
    \label{eq:rho_q_general}
    \\[0.5ex]
    P_q
    &=
    g\int \frac{d^3p}{(2\pi)^3}\,
    \frac{p^2}{3E(p)}\,f_q(E).
    \label{eq:P_q_general}
\end{align}
These expressions reduce to the standard Boltzmann-Gibbs results in the limit $q\to 1$. Since the deformation enters through $f_q(E)$, Eqs.~\eqref{eq:rho_q_general} and \eqref{eq:P_q_general} constitute the basic input for computing the modified radiation and entropy content during the electron-positron annihilation epoch \cite{LesgourguesManganoMielePastor2013,Tsallis2009Book}. For the electron-positron plasma, it is convenient to write the thermodynamic integrals in dimensionless form. Defining
\begin{equation}
    x_e \equiv \frac{m_e}{T},
    \qquad
    y \equiv \frac{p}{T},
    \qquad
    \varepsilon \equiv \frac{E}{T}
    =
    \sqrt{x_e^2+y^2},
    \label{eq:dimensionless_variables_epem}
\end{equation}
where $m_e \approx 0.510~{\rm MeV}$ is the electron mass, and taking $\mu_e\simeq 0$, the deformed energy density is
\begin{align}
   \rho_{e^+e^-,q}(T) &= \frac{g_{e^{+}e^{-}}T^4}{2\pi^2}\nonumber\\ \times &\int_0^{y_{\rm max}} \frac{\varepsilon y^2 dy}{(1+(q-1)\varepsilon)^{1/(q-1)}+1}.
    \label{eq:rho_q_epem_dimensionless}
\end{align}
Similarly, the pressure is given by
\begin{align}
    P_{e^+e^-,q}(T) &= \frac{g_{e^{+}e^{-}}T^4}{6\pi^2}\nonumber\\
    &\times \int_0^{y_{\rm max}} \frac{y^4 dy}{\varepsilon[(1+(q-1)\varepsilon)^{1/(q-1)}+1]}.
    \label{eq:P_q_epem_dimensionless}
\end{align}
Here $g_{e^\pm}=4$ accounts for the two spin states of electrons and positrons. The upper limit is fixed by the domain of the $q$-exponential. With the convention of Eq.~\eqref{eq:q_exponential}, one has
\begin{equation}
    y_{\rm max}
    =
    \begin{cases}
    \displaystyle
    \sqrt{\frac{1}{(1-q)^2}-x_e^2},
    & q<1, \\[2ex]
    \infty,
    & q\geq 1,
    \end{cases}
    \label{eq:ymax_epem}
\end{equation}
provided that the square root is real in the compact-support case. Thus, for $q<1$ the distribution has a finite kinematic domain, while for $q\geq 1$ the integral extends to infinity. These expressions reduce to the standard Fermi-Dirac electron-positron integrals in the limit $q\to 1$ and provide the direct input for the entropy and temperature-ratio analysis during $e^-e^+$ annihilation.

\section{Entropy density and entropic degrees of freedom}
\label{sec:entropy_degrees_freedom}

Using the thermodynamic integrals in Eqs.~\eqref{eq:rho_q_epem_dimensionless} and \eqref{eq:P_q_epem_dimensionless}, we define an electromagnetic entropy density to be used in the instantaneous decoupling estimate:
\begin{align}
    s_{{\rm EM},q}(T)
    &=
    s_{\gamma}(T)
    +
    s_{e^{+}e^{-},q}(T).
    \label{eq:s_EM_split}
\end{align}
This definition should be understood as an effective thermodynamic measure of the electromagnetic entropy budget in the present approximation. In the Curado-Tsallis prescription adopted in this work, the generalized equilibrium distributions are obtained by maximizing the Tsallis entropy under the corresponding macroscopic constraints, and the thermodynamic Legendre structure can be consistently formulated in terms of the conjugate thermodynamic variables \cite{CuradoTsallis1991,TsallisMendesPlastino1998,Tsallis2009Book}. Therefore, within this thermodynamically consistent implementation, we reconstruct the electron-positron entropy contribution from the deformed energy density and pressure through a Gibbs-like relation.

In this work photons are kept extensive, so their entropy density is given by the standard expression
\begin{align}
    s_{\gamma}(T)
    &=
    \frac{2\pi^2}{45}
    g_{\gamma,*s}T^3
    \nonumber\\
    &=
    \frac{4\pi^2}{45}T^3,
    \qquad
    g_{\gamma,*s}=2 .
    \label{eq:s_gamma_standard}
\end{align}
For the electron-positron component, and neglecting the chemical potential, the Gibbs-like relation gives
\begin{align}
    s_{e^{+}e^{-},q}(T)
    &=
    \frac{\rho_{e^+e^-,q}(T)+P_{e^+e^-,q}(T)}{T}.
    \label{eq:s_epem_rho_P}
\end{align}
Here the use of Eq.~\eqref{eq:s_epem_rho_P} does not imply that we derive a full microscopic nonextensive entropy functional for the coupled electromagnetic-neutrino plasma. Rather, it provides the entropy density associated with the deformed equation of state of the electron-positron component, under the assumptions of vanishing chemical potential and preserved thermodynamic Legendre structure. This is the quantity entering the entropy conservation estimate used below.
Substituting Eqs.~\eqref{eq:rho_q_epem_dimensionless} and \eqref{eq:P_q_epem_dimensionless}, one obtains
\begin{align}
    s_{e^{+}e^{-},q}(T)
    &=
    \frac{g_{e^{+}e^{-}}T^3}{2\pi^2}
    \int_0^{y_{\rm max}}
    dy\,
    \nonumber\\
    &\quad\times
    \frac{
        y^2\varepsilon
        +
        \dfrac{y^4}{3\varepsilon}
    }{
        \left[1+(q-1)\varepsilon\right]^{1/(q-1)}
        +1
    } .
    \label{eq:s_epem_q}
\end{align}
Here the same dimensionless variables introduced above are used,
\begin{equation}
    x_e=\frac{m_e}{T},
    \qquad
    y=\frac{p}{T},
    \qquad
    \varepsilon=\sqrt{x_e^2+y^2}.
    \label{eq:dimensionless_variables_entropy}
\end{equation}
Therefore, the total electromagnetic entropy density during the epoch of interest is
\begin{align}
    s_{{\rm EM},q}(T)
    &=
    T^3
    \bigg[
    \frac{4\pi^2}{45}
    +
    \frac{g_{e^{+}e^{-}}}{2\pi^2}
    \int_0^{y_{\rm max}}
    dy\,
    \nonumber\\
    &\quad\times
    \frac{
        y^2\varepsilon
        +
        \dfrac{y^4}{3\varepsilon}
    }{
        \left[1+(q-1)\varepsilon\right]^{1/(q-1)}
        +1
    }
    \bigg].
    \label{eq:s_EM_q}
\end{align}

The corresponding electromagnetic entropic degrees of freedom are defined by
\begin{align}
    g_{*s,{\rm EM},q}(T)
    &\equiv
    \frac{45}{2\pi^2}
    \frac{s_{{\rm EM},q}(T)}{T^3}.
    \label{eq:gstar_s_EM_q_def}
\end{align}
Using Eq.~\eqref{eq:s_EM_q} this gives one of the main tools of this work, the entropic degrees of freedom in the proposed Tsallis scenario:
\begin{align}
    g_{*s,{\rm EM},q}(T)
    &=
    2
    +
    \frac{45\,g_{e^{+}e^{-}}}{4\pi^4}
    \int_0^{y_{\rm max}}
    dy\,
    \nonumber\\
    &\quad\times
    \frac{
        y^2\varepsilon
        +
        \dfrac{y^4}{3\varepsilon}
    }{
        \left[1+(q-1)\varepsilon\right]^{1/(q-1)}
        +1
    } .
    \label{eq:gstar_s_EM_q}
\end{align}
In the extensive limit $q\to 1$, the standard electromagnetic entropy density and entropic degrees of freedom are recovered. Equation~\eqref{eq:gstar_s_EM_q} provides the input for the modified temperature ratio between neutrinos and photons, encoding the Tsallis statistical correction to photon reheating during electron-positron annihilation.

\begin{figure}[!t]
    \centering
    \includegraphics[width=1\linewidth]{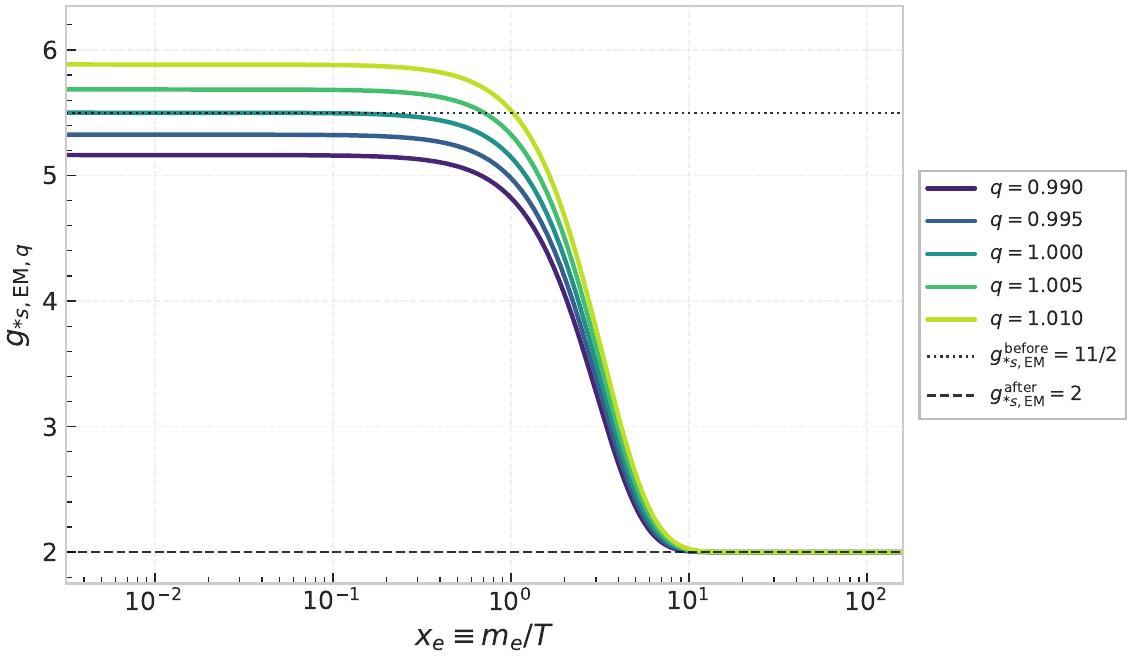}
    \caption{\textbf{Tsallis deformed electromagnetic entropic degrees of freedom.}
    Evolution of $g_{*s,{\rm EM},q}$ as a function of
    $x_e=m_e/T$ for representative values of the nonextensive parameter $q$.
    The standard limits $g_{*s}^{\rm EM,before}=11/2$ and
    $g_{*s}^{\rm EM,after}=2$ are shown as references.}
    \label{fig:gstar_s_EM_q}
\end{figure}

In Fig.~\ref{fig:gstar_s_EM_q} we show the behavior of
Eq.~\eqref{eq:gstar_s_EM_q} as a function of
$x_e=m_e/T$ for different values of the nonextensive parameter $q$.
The standard extensive case, $q=1$, interpolates between the expected
relativistic limit,
$g_{*s}^{\rm EM}=11/2$, valid when $T\gg m_e$, and the late time photon-only
limit, $g_{*s}^{\rm EM}=2$, reached when $T\ll m_e$. The Tsallis deformation
modifies the electron-positron contribution to the electromagnetic entropy:
values of $q>1$ enhance the high energy tail of the distribution and increase
the effective entropic contribution of the pairs, whereas values of $q<1$
suppress it through the compact-support behavior of the $q$-exponential.
Consequently, the transition around $x_e\sim 1$, where electron-positron
annihilation becomes relevant, is shifted with respect to the Boltzmann-Gibbs
case. This temperature-dependent change in $g_{*s,{\rm EM},q}$ is the
thermodynamic origin of the modified photon reheating studied below.

\section{Modified temperature ratio}
\label{sec:modified_temperature_ratio}

As shown in Sec.~\ref{sec:thermal_history}, the standard temperature ratio between neutrinos and photons follows from the conservation of comoving entropy in the electromagnetic sector under the instantaneous decoupling approximation, leading to Eq.~\eqref{eq:Tnu_Tgamma_standard}. In the present framework, the same thermodynamic argument is applied after replacing the standard electromagnetic entropic degrees of freedom by the Tsallis deformed quantity derived in Eq.~\eqref{eq:gstar_s_EM_q}. This allows us to encode the effect of a nonextensive electron-positron plasma on the entropy transfer to photons during annihilation, while keeping the neutrino sector decoupled after a certain decoupling temperature $T_{\rm dec}$. Therefore, the quantity derived below should be understood as an effective temperature ratio within the instantaneous decoupling approximation, not as the result of a kinetic neutrino decoupling calculation.
\begin{equation}
    \left(\frac{T_{\nu}}{T_{\gamma}}\right)_q
    =
    \left[
    \frac{
    g_{*s,{\rm EM},q}(T_{\rm after})
    }{
    g_{*s,{\rm EM},q}(T_{\rm before})
    }
    \right]^{1/3}.
    \label{eq:Tnu_Tgamma_q_general}
\end{equation}
After electron-positron annihilation, the electromagnetic bath is composed
only of photons. Since the photon sector is kept extensive in the present
analysis, one has
\begin{equation}
    g_{*s,{\rm EM},q}(T_{\rm after})=2.
    \label{eq:gstar_after_q}
\end{equation}
Therefore, the deformed neutrino to photon temperature ratio becomes
\begin{equation}
    \left(\frac{T_{\nu}}{T_{\gamma}}\right)_q
    =
    \left[
    \frac{2}
    {g_{*s,{\rm EM},q}(T_{\rm before})}
    \right]^{1/3}.
    \label{eq:Tnu_Tgamma_q}
\end{equation}
In the extensive limit, $q\to1$, Eq.~\eqref{eq:Tnu_Tgamma_q} recovers the
standard photon-reheating result,
$(T_\nu/T_\gamma)_{\rm std}=(4/11)^{1/3}$, discussed in
Sec.~\ref{sec:thermal_history} \cite{KolbTurner1990,LesgourguesManganoMielePastor2013}.
To isolate the relative effect of the Tsallis deformation, we define the
normalized temperature ratio
\begin{equation}
    \mathcal{R}_T(q)
    \equiv
    \frac{(T_\nu/T_\gamma)_q}{(T_\nu/T_\gamma)_{q=1}}.
    \label{eq:RT_q}
\end{equation}
By construction, $\mathcal{R}_T(1)=1$.

\begin{figure}[!t]
    \centering
    \includegraphics[width=1\linewidth]{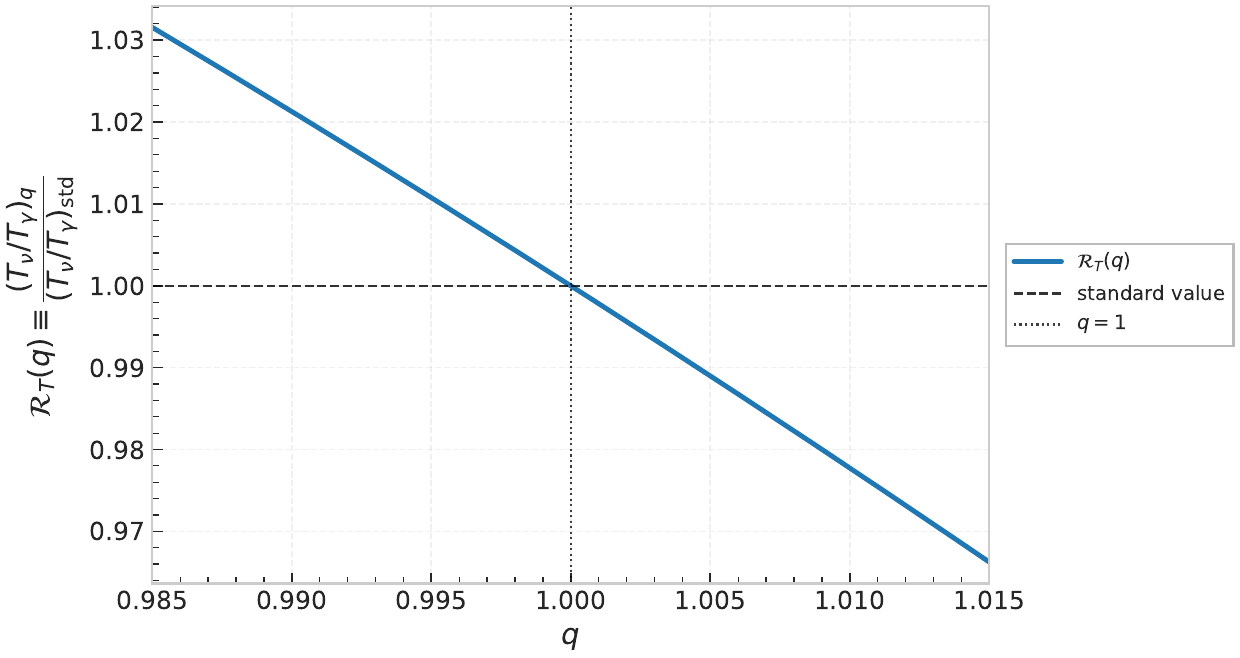}
    \caption{\textbf{Modified neutrino to photon temperature ratio.}
    Relative temperature ratio
    $\mathcal{R}_T(q)=(T_\nu/T_\gamma)_q/(T_\nu/T_\gamma)_{q=1}$
    induced by the Tsallis deformed electromagnetic entropy transfer during
    electron-positron annihilation. The horizontal dashed line denotes the
    standard extensive value, while the vertical dotted line marks the
    Boltzmann-Gibbs limit $q=1$.}
    \label{fig:temperature_ratio_RT_q}
\end{figure}

Figure~\ref{fig:temperature_ratio_RT_q} shows that the Tsallis deformation
induces a monotonic correction to the neutrino to photon temperature ratio. For $q<1$, the compact-support behavior of the $q$-exponential reduces the electron-positron entropic contribution before annihilation, leading to a larger relative value of $(T_\nu/T_\gamma)_q$. Conversely, for $q>1$, the enhanced high energy tail increases the electromagnetic entropy stored in the electron-positron component, producing a stronger photon reheating and
therefore a smaller value of $(T_\nu/T_\gamma)_q$. This normalized ratio is the quantity that will be propagated into the effective radiation content in
the next section.

\section{Mapping to the effective number of relativistic species}
\label{sec:mapping_neff}

Having obtained the modified neutrino to photon temperature ratio in Eq.~\eqref{eq:Tnu_Tgamma_q}, we now map this thermodynamic correction into an effective value of \(N_{\rm eff}\). This mapping is intended to quantify the shift induced by the modified temperature ratio and should not be interpreted as a replacement for a full kinetic computation of \(N_{\rm eff}\).
\begin{equation}
    \rho_r
    =
    \rho_\gamma
    \left[
    1+
    \frac{7}{8}
    \left(\frac{T_\nu}{T_\gamma}\right)^4
    N_{\rm eff}
    \right],
    \qquad
    \rho_\gamma=\frac{\pi^2}{15}T_\gamma^4 .
    \label{eq:rho_r_neff_mapping}
\end{equation}
This expression makes direct and explicit that any modification of the ratio $T_\nu/T_\gamma$ changes the neutrino contribution to the total radiation density through the fourth power of the ratio. In the standard thermal history, the instantaneous decoupling result $(T_\nu/T_\gamma)_{\rm std}=(4/11)^{1/3}$ is corrected by non-instantaneous neutrino decoupling and finite-temperature QED effects, leading to the Standard Model prediction
\begin{equation}
    N_{\rm eff}^{\rm std}\simeq 3.044 .
    \label{eq:Neff_std_value}
\end{equation}
This value will be used as the normalization of the extensive limit
\cite{DeSalasPastor2016,AkitaYamaguchi2020,FrousteyPitrouVolpe2020,BennettEtAl2021Neff}.

In the present framework, the neutrino sector is not directly deformed.
Instead, the Tsallis correction enters through the electromagnetic entropy transfer during electron-positron annihilation, which modifies the final
temperature ratio between neutrinos and photons. Therefore, the corresponding effective number of relativistic species is:
\begin{equation}
    N_{\rm eff}(q)
    =
    N_{\rm eff}^{\rm std}
    \left[
    \frac{
    (T_\nu/T_\gamma)_q
    }{
    (T_\nu/T_\gamma)_{q=1}
    }
    \right]^4 .
    \label{eq:Neff_q_ratio}
\end{equation}
Using the normalized temperature ratio introduced in Eq.~\eqref{eq:RT_q},
this can be written compactly as
\begin{equation}
    N_{\rm eff}(q)
    =
    N_{\rm eff}^{\rm std}\,
    \mathcal{R}_T^4(q).
    \label{eq:Neff_q_RT}
\end{equation}
By construction the extensive limits are recovered,
\begin{equation}
    \lim_{q\to 1} \mathcal{R}_T(q)=1,
    \qquad
    \lim_{q\to 1} N_{\rm eff}(q)=N_{\rm eff}^{\rm std}.
    \label{eq:Neff_extensive_limit}
\end{equation}
This normalization ensures that the standard limit is recovered at \(q=1\). For \(q\neq1\), however, the correction is implemented only through the effective thermodynamic temperature ratio. We do not recompute the non-instantaneous neutrino decoupling contribution in the deformed plasma.

It is useful to define the induced shift with respect to the Standard Model prediction,
\begin{equation}
    \Delta N_{\rm eff}(q)
    \equiv
    N_{\rm eff}(q)-N_{\rm eff}^{\rm std}.
    \label{eq:Delta_Neff_q}
\end{equation}
Substituting Eq.~\eqref{eq:Neff_q_RT}, one obtains
\begin{equation}
    \Delta N_{\rm eff}(q)
    =
    N_{\rm eff}^{\rm std}
    \left[
    \mathcal{R}_T^4(q)-1
    \right].
    \label{eq:Delta_Neff_q_RT}
\end{equation}
Equations~\eqref{eq:Neff_q_RT} and \eqref{eq:Delta_Neff_q_RT} provide the
direct effective mapping between the microscopic nonextensive parameter $q$, which
modifies the electron-positron entropy contribution, and the macroscopic
radiation content measured through $N_{\rm eff}$.
\begin{table}[!t]
\centering
\caption{\textbf{Observational inputs used in the phenomenological comparison.}
The effective thermodynamic estimate \(N_{\rm eff}(q)\) is compared with CMB$+$BAO as the main reference, while the BBN-inferred value is used only as an illustrative Gaussian comparison.}
\label{tab:likelihood_observables}
\small
\setlength{\tabcolsep}{3pt}
\renewcommand{\arraystretch}{1.28}

\begin{tabularx}{\linewidth}{@{}%
>{\raggedright\arraybackslash}X
>{\centering\arraybackslash}p{2.45cm}
>{\centering\arraybackslash}p{2.15cm}
>{\raggedright\arraybackslash}X
@{}}
\toprule
Observable & Symbol & Value $(1\sigma)$ & Reference \\
\midrule
Effective number of relativistic species from CMB$+$BAO
&
$N_{\rm eff}^{\rm CMB+BAO}$
&
$2.99\pm0.17$
&
Planck 2018$+$BAO \cite{Planck2018CosmoParams}
\\
BBN-inferred effective number of relativistic species
&
$N_{\rm eff}^{\rm BBN}$
&
$2.88\pm0.16$
&
BBN primordial abundances \cite{CyburtFieldsOliveYeh2016}
\\
\bottomrule
\end{tabularx}
\end{table}
For small departures from extensivity, this mapping can be understood
perturbatively. If $\mathcal{R}_T(q)=1+\delta_T(q)$ with
$|\delta_T(q)|\ll 1$, then
\begin{equation}
    N_{\rm eff}(q)
    \simeq
    N_{\rm eff}^{\rm std}
    \left[
    1+4\delta_T(q)
    \right],
    \label{eq:Neff_linearized}
\end{equation}
and therefore
\begin{equation}
    \Delta N_{\rm eff}(q)
    \simeq
    4N_{\rm eff}^{\rm std}\delta_T(q).
    \label{eq:Delta_Neff_linearized}
\end{equation}
Thus, even a percent-level modification of the temperature ratio can be
amplified in the radiation density because the neutrino contribution scales
as $(T_\nu/T_\gamma)^4$. In the present scenario, values of $q<1$ reduce the electron-positron
entropic contribution before annihilation and lead to
$\mathcal{R}_T(q)>1$, which implies $\Delta N_{\rm eff}(q)>0$.
Conversely, values of $q>1$ enhance the electromagnetic entropy stored in
the electron-positron plasma, increase photon reheating, and lead to
$\mathcal{R}_T(q)<1$, implying $\Delta N_{\rm eff}(q)<0$. This sign behavior
follows directly from the entropy transfer mechanism discussed in
Sec.~\ref{sec:modified_temperature_ratio}.

\begin{figure}[!t]
    \centering
    \includegraphics[width=1\linewidth]{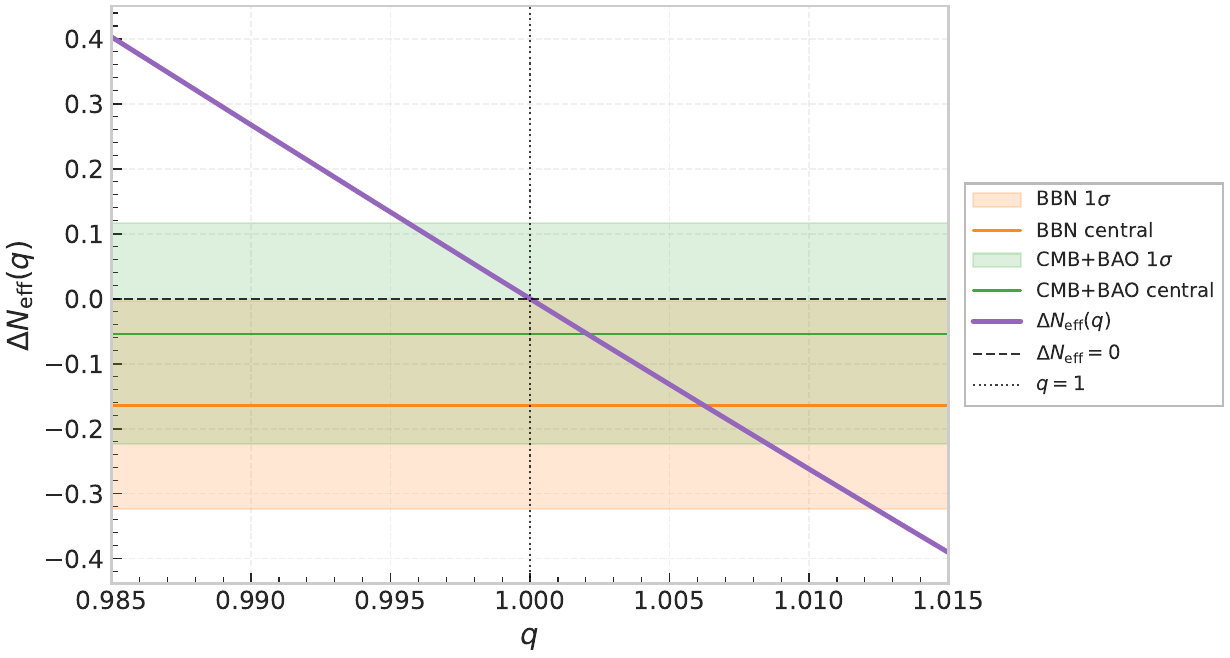}
    \caption{\textbf{Induced shift in the effective number of relativistic species.}
    The curve shows \(\Delta N_{\rm eff}(q)\) obtained by propagating the modified instantaneous decoupling temperature ratio into
    \(N_{\rm eff}(q)=N_{\rm eff}^{\rm std}\mathcal{R}_T^4(q)\).
    The horizontal dashed line denotes the Standard Model limit
    $\Delta N_{\rm eff}=0$, while the vertical dotted line marks the
    Boltzmann-Gibbs value $q=1$. The shaded bands show representative
    observational $1\sigma$ intervals used for comparison in the likelihood
    analysis.}
    \label{fig:delta_Neff_q}
\end{figure}

Figure~\ref{fig:delta_Neff_q} shows the resulting shift
$\Delta N_{\rm eff}(q)$. The curve crosses the Standard Model value at
$q=1$, as required by Eq.~\eqref{eq:Neff_extensive_limit}. For $q<1$, the
modified reheating gives a larger neutrino to photon temperature ratio and therefore a positive contribution to $\Delta N_{\rm eff}$. For $q>1$, the enhanced photon reheating lowers the relative neutrino temperature and produces a negative shift in $N_{\rm eff}$. This function will be used in the next section to perform a phenomenological comparison with observationally inferred values of \(N_{\rm eff}\), using CMB$+$BAO as the main reference and BBN only as an illustrative Gaussian comparison (see Table~\ref{tab:likelihood_observables}) \cite{Planck2018CosmoParams,CyburtFieldsOliveYeh2016}.

\section{Likelihood analysis and results}
\label{sec:likelihood_results}

The likelihood analysis is constructed as a phenomenological comparison between the effective thermodynamic estimate
$N_{\rm eff}(q)$ and the observational determinations summarized in Table~\ref{tab:likelihood_observables}. Since the Tsallis deformation modifies the electromagnetic entropy transfer during electron-positron annihilation, its effect is propagated into the radiation sector through the effective mapping $N_{\rm eff}(q)$ derived in Sec.~\ref{sec:mapping_neff}. We then report the corresponding best-fit values of the nonextensive parameter, \(q_{\rm best}\), by minimizing a Gaussian $\chi^2$ function built from the CMB$+$BAO determination of $N_{\rm eff}$, while the BBN input is included only as an illustrative Gaussian comparison. The statistical construction is defined as
\begin{align}
    &\chi^2_{\rm ill}(q)
    =
    \chi^2_{\rm BBN,ill}(q)
    +
    \chi^2_{\rm CMB+BAO}(q)
    \nonumber\\
    &=
    \frac{
    \left[
    N_{\rm eff}(q)-N_{\rm eff}^{\rm BBN}
    \right]^2
    }{
    \sigma_{\rm BBN}^2
    }
    +
    \frac{
    \left[
    N_{\rm eff}(q)-N_{\rm eff}^{\rm CMB+BAO}
    \right]^2
    }{
    \sigma_{\rm CMB+BAO}^2
    }.
    \label{eq:chi2_q}
\end{align}
where $N_{\rm eff}(q)$ is the effective thermodynamic estimate obtained from
Eq.~\eqref{eq:Neff_q_RT}, while $N_{\rm eff}^{\rm BBN}$ and
$N_{\rm eff}^{\rm CMB+BAO}$ are the central observational values from BBN and CMB$+$BAO, respectively. The quantities $\sigma_{\rm BBN}$ and $\sigma_{\rm CMB+BAO}$ denote the corresponding
$1\sigma$ observational uncertainties. The individual CMB$+$BAO and BBN illustrative profiles are obtained by retaining the corresponding term in Eq.~\eqref{eq:chi2_q}, while \(\chi^2_{\rm ill}(q)\) denotes their illustrative Gaussian combination. The BBN term should not be interpreted as a full primordial-abundance likelihood, since such a treatment would require propagating the modified thermal history into ${\rm D/H}$, $Y_p$, and the neutron-proton conversion rates. To display the phenomenological interval for the nonextensive parameter, we use the shifted
profile
\begin{equation}
    \Delta\chi^2(q)
    =
    \chi^2_{\rm ill}(q)-\chi^2_{\rm ill,min}.
    \label{eq:delta_chi2_q}
\end{equation}

\begin{figure}[!t]
    \centering
    \includegraphics[width=1\linewidth]{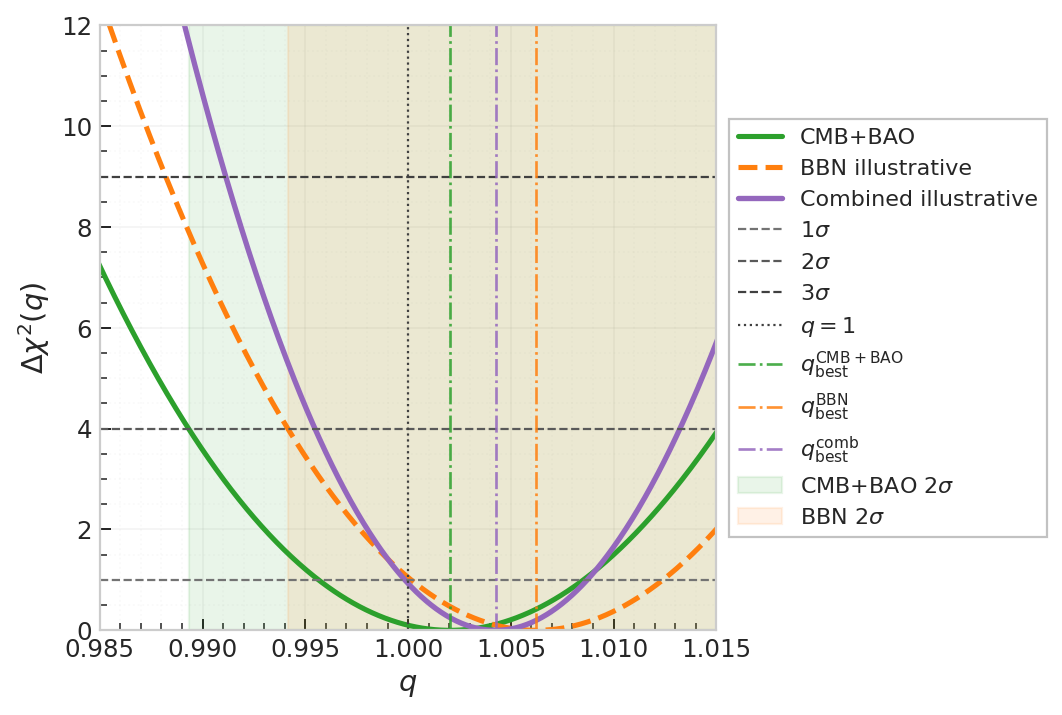}
    \caption{\textbf{Phenomenological profiles for the Tsallis parameter.}
    Shifted Gaussian profiles $\Delta\chi^2(q)$ obtained from the CMB$+$BAO determination, from the BBN-inferred value of $N_{\rm eff}$, and from their illustrative Gaussian combination. The CMB$+$BAO profile is used as the main quantitative reference, while the BBN profile is shown only as an illustrative comparison. The vertical dotted line marks the Boltzmann-Gibbs limit $q=1$.}
    \label{fig:delta_chi2_q}
\end{figure}

\begin{table}[!t]
\centering
\caption{\textbf{Phenomenological intervals for $q$.}
The CMB$+$BAO entry is the main reference; BBN and combined entries are
illustrative Gaussian comparisons.}
\label{tab:q_intervals}
\footnotesize
\setlength{\tabcolsep}{2.5pt}
\renewcommand{\arraystretch}{1.15}
\begin{tabular}{lccc}
\toprule
Input & $q_{\rm best}$ & $2\sigma$ range & $2\sigma$ estimate \\
\midrule
CMB$+$BAO
& $1.0020$
& $0.9893$--$1.0151$
& $|q-1|\leq1.51\times10^{-2}$
\\
BBN ill.
& $1.0062$
& $0.9941$--$1.0187$
& $|q-1|\leq1.87\times10^{-2}$
\\
Comb. ill.
& $1.0043$
& $0.9955$--$1.0132$
& $|q-1|\leq1.32\times10^{-2}$
\\
\bottomrule
\end{tabular}
\end{table}

This quantity removes the absolute normalization of the Gaussian profile and allows
one to identify indicative confidence regions directly from the increase of
$\chi^2_{\rm ill}$ with respect to the best fit, as shown in Fig.~\ref{fig:delta_chi2_q}, for the three cases: CMB$+$BAO, BBN in its illustrative form, and their illustrative Gaussian combination.

This estimate indicates that, within the instantaneous decoupling approximation, any large nonextensive correction to the electromagnetic
entropy transfer produced by electron-positron pair annihilation would be disfavored by the adopted observational intervals. The interval obtained here should be interpreted as a minimal thermodynamic estimate derived
from the phenomenological effect of the Tsallis deformation on photon reheating. A full BBN constraint would require extending the analysis to primordial abundance observables, in particular ${\rm D/H}$ and $Y_p$, which are sensitive to the expansion rate and to the radiation content during BBN \cite{CyburtFieldsOliveYeh2016,PitrouCocUzanVangioni2018}. This would
require propagating the Tsallis deformed thermal history into a full BBN scenario, following standard numerical treatments of primordial nucleosynthesis
\cite{PisantiEtAl2008,PisantiManganoMieleMazzella2021}. A complementary improvement would be to go beyond the entropy-conservation
approximation adopted in this work and solve the relevant Boltzmann equations with collision terms, as done in precision studies of neutrino decoupling and
electron-positron annihilation \cite{ManganoEtAl2005,AkitaYamaguchi2020,FrousteyPitrouVolpe2020,BennettEtAl2021Neff,ThomasDezenGrohsKishimoto2020}.
Such an extension would allow one to test whether the thermodynamic correction derived here remains stable once the non-instantaneous and kinetic aspects of the plasma are included.

\section{Conclusions}
\label{sec:conclusions}

In this work we studied a minimal Tsallis statistical correction to the
entropy transfer produced by electron-positron annihilation in the early Universe. The deformation was introduced through generalized
electron-positron distribution functions, while the photon and neutrino sectors were kept standard. With this prescription, the nonextensive
correction is isolated in the electromagnetic plasma and then propagated into the thermodynamic quantities that determine photon reheating.

The central object of the analysis was the electromagnetic entropy density $s_{{\rm EM},q}(T)$ and the corresponding entropic degrees of freedom $g_{*s,{\rm EM},q}(T)$. Starting from the deformed distribution function $f_q(E)$, we computed the modified energy density and pressure of the electron-positron plasma, $\rho_{e^+e^-,q}(T)$ and $P_{e^+e^-,q}(T)$, and used them to construct the entropy contribution of the electromagnetic
sector. This modification changes the entropy available before
electron-positron annihilation and therefore changes the final temperature ratio between neutrinos and photons.

The modified temperature ratio was then propagated into the effective number of relativistic species through
\begin{equation}
N_{\rm eff}(q)=N_{\rm eff}^{\rm std}\mathcal{R}_T^4(q).
\end{equation}
This normalization preserves the standard result in the Boltzmann-Gibbs limit, since $\mathcal{R}_T(1)=1$ and therefore
$N_{\rm eff}(1)=N_{\rm eff}^{\rm std}$.

The sign of the correction follows from the behavior of the deformed
distribution. For $q>1$, the high energy tail is enhanced, increasing the electron-positron entropic contribution to the electromagnetic plasma. This produces stronger photon reheating, lowers the relative neutrino temperature,
and gives a negative shift in $N_{\rm eff}$. For $q<1$, the compact-support behavior suppresses the electron-positron entropic contribution before
annihilation, reducing photon reheating and increasing the final
neutrino to photon temperature ratio. Therefore, the induced correction is positive for $q<1$ and negative for $q>1$.

Finally, we performed a phenomenological comparison between the effective thermodynamic estimate \(N_{\rm eff}(q)\) and observationally inferred values of \(N_{\rm eff}\). The CMB$+$BAO profile was used as the main quantitative reference, while the BBN and combined profiles were shown as illustrative Gaussian comparisons. This gives an order of magnitude interval for the nonextensive parameter, with the results shown in Table \ref{tab:q_intervals} at the \(2\sigma\) level. Within the instantaneous decoupling approximation, this suggests that sizeable Tsallis corrections to the electromagnetic entropy transfer during electron-positron annihilation must remain close to the Boltzmann-Gibbs limit during the MeV era.

The result obtained here should be interpreted as a minimal thermodynamic benchmark, not as a precision neutrino decoupling or BBN constraint. A full BBN constraint would require including primordial abundance observables, in particular ${\rm D/H}$ and $Y_p$, through a dedicated abundance likelihood. A complementary extension would be to replace the entropy-conservation approximation by a kinetic treatment based on Boltzmann equations with collision terms, allowing non-instantaneous neutrino decoupling, electron-positron annihilation, and finite-temperature plasma effects to be treated in a unified framework.

\section*{Acknowledgements}

MPG. acknowledges Vicerrector\'{\i}a de Investigaci\'on y Desarrollo Tecnol\'ogico (VRIDT) at Universidad Cat\'olica del Norte (UCN) for the scientific support provided by N\'ucleo de Investigaci\'on en Simetr\'{\i}as y la Estructura del Universo (NISEU-UCN), Resoluci\'on VRIDT N$^\circ$200/2025.

\noindent MPG. acknowledges the support and discussions with fellow graduate students at \textit{
Universidad Catolica del Norte.}

\noindent MPG. acknowledges the financial support of the \textit{Direcci\'on general de postgrado}.

\end{document}